\documentclass[nofootinbib,aps,11pt,preprintnumbers,unsortedaddress,prd]{revtex4}
\usepackage{graphicx}
\usepackage{cancel}
\usepackage{amssymb}
\usepackage{textcomp}
\usepackage{amsmath}
\usepackage{bm}
\usepackage{times}
\usepackage{epsfig}
\usepackage{color}
\usepackage{graphics}
\usepackage{hyperref}
\usepackage{setspace}
\usepackage{comment}
\usepackage{epsfig}
\usepackage{amsmath}
\usepackage{bm}
\usepackage{times}
\usepackage{graphicx}
\usepackage{color}
\usepackage{slashed}
\usepackage{graphicx}
\usepackage{amsmath, amssymb}

\hypersetup{
    pdfnewwindow=true,      
    colorlinks=true,       
    linkcolor=black,          
    citecolor=blue,        
    filecolor=blue,      
    urlcolor=blue           
}

\def\bea{\begin{eqnarray}}
\def\eea{\end{eqnarray}}
\begin{document}
\title{Spontaneous $CP$ Violation and the Strong $CP$ Problem in Left-Right Symmetric Theories}
\author{Sebastian Ohmer}
\affiliation{Max-Planck-Institut f\"{u}r Kernphysik, Saupfercheckweg 1, D-69117 Heidelberg, Germany}
\begin{abstract}
We study spontaneous $CP$ violation as a solution to the strong $CP$ problem in left-right symmetric theories. The discrete $CP$ symmetry is broken by a complex vacuum expectation value of a right-handed Higgs doublet. Heavy vectorlike down-type quarks mix with the Standard Model quarks introducing the known $CP$ violation -- realizing a variant of the Nelson-Barr mechanism. A nonvanishing QCD vacuum angle is generated at loop level. The implementation in the ultraviolet complete theory of trinification at low scales is discussed. We further comment on the phenomenology and future testability of the model.
\end{abstract}
\maketitle
\section{Introduction}
The Standard Model of particle physics has been a success story for the last four decades. However, there still remain numerical puzzles such as the hierarchy problem, the proton stability, or the flavor structure of the Standard Model. In this article, we want to address one of these puzzles: the strong $CP$ problem.

The non-Abelian $SU(3)_C$ gauge symmetry of the Standard Model has a nontrivial vacuum structure which allows for explicit $CP$ violation via the QCD vacuum angle $\theta_{\text{QCD}}$~\cite{tHooft1,tHooft2}
\begin{align}
\mathcal{L}\supset \frac{g_s^2 \theta_{\text{QCD}}}{32 \pi^2} \text{Tr}\left( G_{\mu\nu} \tilde{G}^{\mu \nu} \right)\,,
\end{align}
where $g_s$ is the QCD gauge coupling and $G_{\mu \nu}$ is the gluon field strength tensor with $\tilde{G}_{\mu \nu} = (1/2) \epsilon_{\mu \nu \alpha \beta} G^{\alpha \beta}$. Measurements of the neutron electric dipole moment~\cite{EDM_bound} and of the electric dipole moment of mercury~\cite{EDM_mercury1,EDM_mercury2} have set an experimental upper limit on the QCD vacuum angle
\begin{align}
\theta_{\text{QCD}} \lesssim 10^{-10} \,.
\label{eq_lim_theta}
\end{align}
The $CP$ violation in QCD is therefore strikingly small and the strong $CP$ problem arises: Why is the $CP$ violation in the weak interaction so strong compared to the $CP$ violation in the strong interaction?

In principle, there are three known solutions to the strong $CP$ problem.
\begin{itemize}
\item A zero up-quark mass at tree-level. The measured up-quark mass is then only generated via nonperturbative QCD processes. The theory then exhibits an axial symmetry which renders $\theta_{\text{QCD}}$ nonphysical~\cite{Solution_mU0_1,Solution_mU0_2}.
\item Add a new scalar field with a nontrivial charge with respect to an anomalous global $U(1)_{PQ}$ symmetry. After spontaneously breaking $U(1)_{PQ}$ the pseudo-Nambu-Goldstone boson of $U(1)_{PQ}$, the QCD axion, dynamically forces $\theta_{\text{QCD}}$ to be small. This solution was first discussed in Ref.~\cite{PQ1,PQ2} and the $U(1)_{PQ}$ symmetry is thus referred to as Peccei-Quinn symmetry.
\item The $CP$ symmetry is an exact symmetry of nature which is broken spontaneously by a complex vacuum expectation value~\cite{sCPV1,sCPV2,sCPV3,sCPV4}. All the $CP$ violation in the Standard Model is then connected to the complex phase of a vacuum expectation value.
\end{itemize}
Today, vanishing quark masses at tree-level are in conflict with state-of-the-art lattice calculations~\cite{Lattice_Review}. A zero up-quark mass at tree-level as solution to the strong $CP$ problem is thus mostly discarded. The theoretically preferred solution is the QCD axion. Axionlike particles received more and more attention in recent years~\cite{axion_review1,axion_review2} -- especially as there is a significant experimental effort to test the axion hypothesis~\cite{axion_exp,axion_exp2}. Nevertheless, we should not forget that the Peccei-Quinn symmetry also has theoretical short comings such as the stabilization with respect to Planck scale physics~\cite{axion_prob1,axion_prob2,axion_prob3}.

In this paper, we evaluate the theoretical attractiveness of spontaneous $CP$ violation as solution to the strong $CP$ problem. We thereby focus on the Nelson-Barr mechanism. The Nelson-Barr mechanism was first introduced and phenomenologically discussed in Ref.~\cite{sCPV1,sCPV2,sCPV3,sCPV4}, see Ref.~\cite{NB_prob} for a recent review. The mechanism requires additional heavy vectorlike quarks which mix with the Standard Model quarks via a complex coupling. As the determinant of the quark mass matrix has to be real at tree-level in order to allow for a vanishing QCD vacuum angle $\theta_\text{QCD}$ at leading order, the mixing has to be asymmetric. A small nonvanishing vacuum angle $\theta_\text{QCD}$ can then be produced only at loop level. A first bottom-up toy model discussing the Nelson-Barr mechanism had to require additional discrete symmetries to suppress $\theta_{\text{QCD}}$~\cite{BBP_model0,BBP_model}. This is a problem which generally arises in the Nelson-Barr mechanism~\cite{NB_prob}. We therefore consider left-right symmetric theories~\cite{LR1,LR2,LR3,LR4} because we only observe left-handed currents at low energies and thus it is plausible that such theories provide the preferred framework for such an asymmetric symmetry breaking and quark mixing.

The paper is organized as follows. In Sec.~\ref{NBLR}, we introduce the minimal realization of the Nelson-Barr mechanism in left-right symmetric theories. We discuss the embedding of this minimal theory in the UV complete theory of trinification in Sec.~\ref{NBTri}. Section~\ref{Pheno} is then dedicated to the phenomenology of spontaneous $CP$ violation and the possibility to definitively probe these theories in future experiments. Finally, we conclude in Sec.~\ref{con}.

\section{Nelson-Barr Mechanism in Left-Right Symmetry} 
\label{NBLR}
The Nelson-Barr mechanism realizes the idea of spontaneous $CP$ violation as a solution to the strong $CP$ problem. Thereby, new vectorlike quarks are added to the Standard Model of particle physics which mix with the known quarks. The mixing takes place via a new scalar degree of freedom which acquires a complex vacuum expectation value and hence spontaneously breaks the $CP$ symmetry. The complex phase of the vacuum expectation value then allows for a connection of the $CP$ violation in the Cabibbo-Kobayashi-Maskawa (CKM) matrix and the QCD vacuum angle.

An especially appealing scenario for spontaneous $CP$ violation are left-right symmetric theories. It was first realized~\cite{Pconservation1,Pconservation2} that parity is sufficient to solve the strong $CP$ problem in left-right symmetric theories without a scalar bidoublet. The Standard Model fermion masses are then generated by nonrenormalizable dimension five operators. A recent discussion of the dimension five mass terms in the context of Froggatt-Nielsen theories can be found in Ref.~\cite{Hall}. A solution to the strong $CP$ problem without an axion and without a scalar bidoublet but with renormalizable mass terms in the context of left-right symmetric theories was first studied in Ref.~\cite{LR_sCPV1}. By not including a scalar bidoublet in the theory, heavy vectorlike quarks and leptons had to be introduced in order to give mass to the Standard Model fermions via left- and right-handed Higgs doublets.

A scalar bidoublet is the preferred mechanism to generate Standard Model fermion masses in left-right symmetric theories. However, it was argued that without supersymmetry a scalar bidoublet in a left-right symmetric theory cannot have a complex vacuum expectation value~\cite{LR_sCPV2}. Further nonsupersymmetric studies of spontaneous $CP$ violation in left-right symmetric theories can be found in Refs.~\cite{LR_CP1,LR_CP2,LR_CP3,LR_CP4,LRspontaneousCPtriplets,Nierste}. In this paper, we want to address the question: What is the left-right symmetric theory with the minimal number of propagating degrees of freedom which can spontaneously break $CP$ and solve the strong $CP$ problem?

Minimal left-right symmetric gauge theories are based on the gauge symmetry group
\begin{align}
G_\text{LR} =  SU(3)_C \times SU(2)_L \times SU(2)_R \times U(1)_{B-L}\,.
\end{align}
Without supersymmetry and with only new vectorlike down-type quarks to realize the Nelson-Barr mechanism, the minimal fermionic particle content is given by
\begin{align}
&q_L \sim (3, 2, 1, 1/3)\,, \quad & &q_R \sim (3, 1, 2, 1/3)\,, \nonumber \\
&\ell_L \sim (1, 2, 1, -1)\,, \quad & &\ell_R \sim (1, 1, 2, -1)\,, \nonumber \\
&D_L \sim (3, 1, 1, -2/3)\,, \quad & &D_R \sim (3, 1, 1, -2/3	)\,.
\end{align}
Note that the anomaly conditions allow to add just one generation of vectorlike quarks for simplicity. However, in a unified picture we would expect three generations of additional down-type quarks $D$. Furthermore, we add two types of vectorlike fermions per generation less than in Ref.~\cite{LR_sCPV1}. The scalar sector we consider is given by
\begin{align}
\phi \sim (1, 2, 2, 0)\,, \quad  H_{L,i} \sim (1, 2, 1, 1)\,, \quad  H_{R,j} \sim (1, 1, 2, 1)\,,
\end{align}
with $i,j \in \{1,2\}$. Compared to Ref.~\cite{LR_sCPV1} only a scalar bidoublet was added. The interactions are thus given by
\begin{align}
-\mathcal{L} \supset &\,  \bar{q}_L (y_1 \phi + y_2 \tilde{\phi})q_R + \bar{\ell}_L (y_3 \phi + y_4 \tilde{\phi}) \ell_R + m_D \bar{D}_L D_R \nonumber \\
&+ y_{L,i}\, \bar{q}_L H_{L,i} D_R + y_{R,j}\,\bar{q}_R H_{R,j} D_L + \text{H.c.} \,,
\end{align}
with $\tilde{\phi} = \sigma_2 \phi^* \sigma_2$ and implicit summation over $i$ and $j$. The discrete spacetime symmetries $P$, $C$, and $CP$ are conserved for real Yukawa couplings and the additional condition
\begin{align}
y_{L,i} = y_{R,i}^T \,.
\end{align}
Note that in Ref.~\cite{sCPV_Cond} it was demonstrated that at least two right-handed Higgs doublets are required to spontaneously violate $CP$. The two left-handed Higgs doublets do not acquire a vacuum expectation value and are only included to have a left-right symmetric particle content. The scalar potential takes the form
\begin{align}
V =&\, -\mu^2_{H,ij}\left(H^\dagger_{L,i}H_{L,j} + H^\dagger_{R,i}H_{R,j}\right) + \lambda^{(1)}_{H,ijkl} \left(H^\dagger_{L,i}H_{L,j}H^\dagger_{L,k}H_{L,l} + H^\dagger_{R,i}H_{R,j}H^\dagger_{R,k}H_{R,l} \right) \nonumber \\
&\, +\lambda^{(2)}_{H,ij} H^\dagger_{L,i}H_{L,j}H^\dagger_{R,i}H_{R,j} \nonumber - \mu^2_{\phi,ij}\text{Tr}\left(\phi^\dagger_i \phi_j\right) + \lambda^{(1)}_{\phi,ijkl} \text{Tr}\left(\phi^\dagger_{i}\phi_{j}\right)\text{Tr}\left(\phi^\dagger_{k}\phi_{l}\right) \nonumber \\
&\, +\lambda^{(2)}_{\phi,ijkl} \text{Tr}\left(\phi^\dagger_{i}\phi_{j} \phi^\dagger_{k}\phi_{l}\right) + a_{ijkl}\left(H^\dagger_{L,i}H_{L,j} + H^\dagger_{R,i}H_{R,j}\right) \text{Tr}\left(\phi^\dagger_k \phi_l\right) \nonumber \\
&\,+ b_{ijkl} \left(H^\dagger_{L,i}\phi_j \phi^\dagger_k H_{L,l} + H^\dagger_{R,i}\phi_j \phi^\dagger_k H_{R,l} \right)\,,
\end{align}
with $i,j,k,l \in \{1,2\}$, $\phi_1 \equiv \phi$, and $\phi_2 \equiv \tilde{\phi}$. We had to assume a discrete $Z_3$ symmetry 
\begin{align}
\phi_1 \to \textbf{1}'\, \phi_1 \quad \text{and} \quad \phi_2 \to \textbf{1}''\, \phi_2\,,
\end{align}
to forbid linear terms and couplings introducing a complex electroweak symmetry breaking vacuum expectation value. $CP$ conservation only allows for real couplings in the scalar potential. The relevant interactions to generate a complex vacuum expectation value are those with differing spurion charges~\cite{sCPV_Cond}. Assigning $H_{R,1}$ and $H_{R,2}$ the spurion charges
\begin{align}
H_{R,1} \rightarrow e^{ia} H_{R,1} \quad \text{and} \quad H_{R,2} \rightarrow e^{ib} H_{R,2} \,,
\end{align}
we find that the relevant terms in the scalar potential to spontaneously break $CP$ are given by
\begin{align}
V \supset &\, - \mu^2_{H, 12} H^\dagger_{R,1} H_{R,2} + \bigg( \lambda^{(1)}_{H,1212} H^\dagger_{R,1} H_{R,2} H^\dagger_{R,1} H_{R,2} + \lambda^{(1)}_{H,1112} H^\dagger_{R,1} H_{R,1} H^\dagger_{R,1} H_{R,2} \nonumber\\
&\,+ \lambda^{(1)}_{H,2221} H^\dagger_{R,2} H_{R,2} H^\dagger_{R,2} H_{R,1} + \text{H.c.} \bigg) \,,
\end{align}
with spurion charges
\begin{align}
[\mu^2_{H, 12}] = -a + b\,,\quad [\lambda^{(1)}_{H,1212}] = -2 a + 2 b\,, \quad [\lambda^{(1)}_{H,1112}] = -a + b \,, \quad [\lambda^{(1)}_{H,2221}] = a - b\,.
\end{align}

As a proof of principle that such a scalar potential can generate spontaneous $CP$ violation and realize the Nelson-Barr mechanism, we assume that all couplings mixing the bidoublet $\phi$ with the Higgs doublets $H_{L,i}$ and $H_{R,j}$ are negligible,
\begin{align}
a_{ijkl},b_{ijkl} \rightarrow 0 \quad \text{for all  } i,j,k,l\,.
\end{align}
To further simplify the discussion, we take the limit
\begin{align}
\lambda^{(1)}_{H,1112}, \lambda^{(1)}_{H,1122}, \lambda^{(1)}_{H,2221}, \lambda^{(2)}_{H,12} \rightarrow 0 \,,
\end{align}
such that the mixing of the four Higgs doublets is controlled by only four couplings
\begin{align}
\mu^2_{H, 12}, \lambda^{(1)}_{H,1212}, \lambda^{(2)}_{H,11}, \lambda^{(2)}_{H,22}\,.
\end{align}
A minimum of the scalar potential is then given by
\begin{align}
\langle \phi \rangle = \begin{pmatrix} v_1 && 0 \\ 0 && v_2 \end{pmatrix} \,, \quad \langle H_{L,i} \rangle = \begin{pmatrix} 0 \\ 0 \end{pmatrix} \,, \quad \langle H_{R,1} \rangle  = \begin{pmatrix} 0 \\ v_{R1}\, e^{i \alpha} \end{pmatrix} \,, \quad \langle H_{R,2} \rangle = \begin{pmatrix} 0 \\ v_{R2} \end{pmatrix} \,,
\label{eq_minimum}
\end{align}
with
\begin{align}
V =&\,  -\mu^2_{\phi,11} \left(v^2_1 + v^2_2\right) - 2 \mu^2_{\phi,12} v_1 v_2 + \lambda^{(1)}_{\phi} \left( v^4_1 + v^4_2 \right) + \lambda^{(2)}_{\phi} \left( v^3_1 v_2 + v_1 v^3_2 \right) \nonumber \\
&\, - \mu^2_{H, 11} v^2_{R1} - \mu^2_{H, 22} v^2_{R2} - 2 \mu^2_{H, 12} v_{R1} v_{R2} \text{cos}(\alpha) + 2 \lambda^{(1)}_{H, 1212} v^2_{R1} v^2_{R2} \text{cos}(2\alpha) \nonumber \\
&\, + \lambda^{(1)}_{H, 1111} v^4_{R1} + \lambda^{(1)}_{H, 2222} v^4_{R2}\,.
\end{align}
The vacuum expectation values (\ref{eq_minimum}) then are a valid minimum for
\begin{align}
&\, v^2_1 \simeq \frac{\left(16 \left(\lambda^{(1)}_\phi\right)^2 - 3 \left(\lambda^{(2)}_\phi \right)^2 \right) \mu^2_{\phi, 11}}{ 32 \left(\lambda^{(1)}_\phi \right)^3 }\,, \quad \quad \quad \quad \quad
\, v^2_2 \simeq \frac{ \left(\lambda^{(2)}_\phi \right)^3 \mu^2_{\phi, 11}}{32 \left(\lambda^{(1)}_\phi \right)^3 }\,, \nonumber \\
&\, v^2_{R1} \simeq \frac{\mu^2_{H,11}}{2 \lambda^{(1)}_{H, 1111}}\,, \quad \quad
\, v^2_{R2} \simeq \frac{\mu^2_{H,22} }{2 \lambda^{(1)}_{H, 2222} }\,, \quad \quad
\, \text{cos}(\alpha) = \frac{\mu^2_{H,12}}{4 \lambda^{(1)}_{H, 1212}\, v_{R1}\, v_{R2}}\,,
\end{align}
in the limit of $\lambda^{(1)}_\phi \gg \lambda^{(2)}_\phi > 0$ and $\lambda^{(1)}_{H, 1111},\lambda^{(1)}_{H, 2222} \gg \lambda^{(1)}_{H, 1212} > 0$. The vacuum expectation values are all positive and well-defined for positive quartic couplings. To ensure that $\langle H_{L,i} \rangle = (0, 0)^T$ is a valid minimum of the scalar potential, the conditions
\begin{align}
\lambda^{(2)}_{H,11} \ge 4 \lambda^{(1)}_{H,1111}\,,\quad \lambda^{(2)}_{H,22} \ge 4 \lambda^{(1)}_{H,2222}\,, \quad \mu^2_{12} < \mu_1 \mu_2 \,,
\end{align}
have to be satisfied. The potential is therefore bounded and the extremal solutions belong to a minimum of the potential. This completes our proof of principle and demonstrates that spontaneous $CP$ violation with complex vacuum expectation values in the right-handed Higgs sector and vanishing vacuum expectation values in the left-handed Higgs sector in left-right symmetric theories can be realized.

The scalar bidoublet $\phi$ gives mass to the Standard Model fermions. The lepton and up-type quark mass matrices are given by
\begin{align}
m_\nu = y_3 v_1 + y_4 v_2\,, \quad m_e = y_3 v_2 + y_4 v_1\,, \quad m_u = y_1 v_1 + y_2 v_2 \,,
\end{align}
with implicit flavor indices. We will work in the basis where $m_u$ is diagonal. Note that neutrinos are Dirac fermions in this minimal configuration. Majorana neutrinos could be accounted for by introducing a singly charged scalar field~\cite{minLR}. The down-type quark mass matrix is given by
\begin{align}
M_d = \begin{pmatrix} m_d && 0 \\ m_{dD} && m_D  \end{pmatrix} \,,
\end{align}
in the down-type quark basis $(d, D)$ and with
\begin{align}
m_d = y_1 v_2 + y_2 v_1 \,, \quad m_{dD} = y^T_{R1} v_{R1} e^{-i\alpha} + y^T_{R2} v_{R2}\,,
\end{align}
where flavor indices are again implicit. We can then define a biunitary transformation
\begin{align}
U_L^\dagger M_d U_R = \mathcal{M}_d = \text{diag}\left(m_1, m_2, m_3, m_4\right)\,,
\end{align}
where we parametrize the transformation as
\begin{align}
U_L = \begin{pmatrix} C && D \\ E && F \end{pmatrix}\,,
\end{align}
with $C$ the known $3 \times 3$ CKM matrix
\begin{align}
C = \begin{pmatrix} 1 && 0 && 0 \\ 0 && c_{23} && s_{23} \\ 0 && -s_{23} && c_{23} \end{pmatrix}
\begin{pmatrix} c_{13} && 0 && s_{13}e^{-i \delta} \\ 0 && 1 && 0 \\ -s_{13}e^{i \delta} && 0 && c_{13} \end{pmatrix}
\begin{pmatrix} c_{12} && s_{12} && 0 \\ -s_{12} && c_{12} && 0 \\ 0 && 0 && 1 \end{pmatrix}\,,
\end{align}
with the abbreviations $s_{ij} = \text{sin}(\theta_{ij})$ and $c_{ij} = \text{cos}(\theta_{ij})$ and the measured values $\theta_{12} = (13.04 \pm 0.05)^\circ$, $\theta_{13} = (0.201 \pm 0.011)^\circ$, $\theta_{23} = (2.38 \pm 0.06)^\circ$, and $\delta = (1.20 \pm 0.08)\text{rad}$~\cite{PDG}. Using the relation~\cite{BBP_model}
\begin{align}
C \mathcal{M}^2_d C^{-1} = m_d m_d^\dagger - m_d m_{dD}^\dagger m_{dD} m_d^\dagger \left(m_{dD} m_{dD}^\dagger + m_D^2 \right)^{-1}\,,
\end{align}
we can establish a direct connection between the measured $CP$ violation in the CKM matrix and the $CP$ violating phase $\alpha$. A direct correspondence for example
\begin{align}
\delta = - \alpha\,,
\label{eq_relphase}
\end{align}
is achieved in the limit
\begin{align}
m_{11} = -  \frac{m_{12}m_{D2}}{m_{D1}} \,, \quad m_{d1} = -  \frac{m_{d2}m_{D2}}{m_{D1}} \,, \quad m_{D3} = 0 \,,
\end{align}
where we parametrized the full flavor matrices as
\begin{align}
m_d = \begin{pmatrix} m_{11} && m_{12} && m_{13} \\ m_{12} && m_{22} && m_{23} \\ m_{13} && m_{23} && m_{33}  \end{pmatrix} \,,
\end{align}
and
\begin{align}
m_{dD} = \left(m_{D1}e^{-i \alpha} + m_{d1}, m_{D2}e^{-i \alpha} + m_{d2}, m_{D3}e^{-i \alpha} + m_{d3} \right) \,.
\end{align}

\begin{figure}[!htb]
	\centering
		\includegraphics[width=0.45\textwidth]{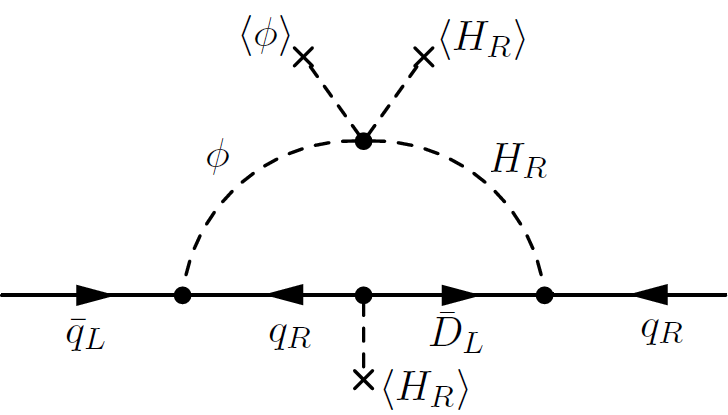}
	\caption{One-loop diagram contributing to the QCD vacuum angle $\theta_{\text{QCD}}$.}
	\label{fig_1loop}
\end{figure}

Due to the asymmetric mixing in the down-type quark mass matrix, we find at tree-level
\begin{align}
\text{Arg}\left(\text{det}\left( M_d \right) \right) = 0 \,.
\end{align}
Thus, as $CP$ is an exact symmetry of the Lagrangian the QCD vacuum angle $\theta_{\text{QCD}}$ vanishes at tree-level. This changes at the quantum level. One-loop processes as shown for example in Fig.~\ref{fig_1loop} contribute to $m_d$ with the result that $\theta_{\text{QCD}}$ is no longer vanishing. As these corrections are not only loop suppressed but also Yukawa suppressed, the resulting $\theta_{\text{QCD}}$ is compatible with current constraints
\begin{align}
\Delta \theta_{\text{QCD}} \simeq 10^{-11} \left(\frac{y_{1/2}}{0.01}\right) \left(\frac{y_R}{0.01}\right)^2 \bigg(\frac{a}{0.01}\bigg)\,.
\end{align}
The theory is thus also testable with future electric dipole moment measurements of the neutron~\cite{nEDM1,nEDM2,nEDM3}.

The first higher dimensional operator contributing to $m_D$ is given by
\begin{align}
\mathcal{L} \supset \frac{H^\dagger_{R,1} H_{R,2} \bar{D}_L D_R}{\Lambda} + 1 \leftrightarrow 2 \,,
\label{eq_dim5Op}
\end{align}
which results in a mass correction
\begin{align}
\Delta m_D \simeq \frac{v_{R1}v_{R2}\,e^{i\alpha}}{\Lambda} \,,
\end{align}
assuming order one couplings. If we further assume $v_R \equiv v_{R1} \simeq v_{R2}$, we find for a Planck scale suppressed operator the upper bound~\cite{NB_prob}
\begin{align}
v_R \lesssim 10^8 \text{ GeV}\,.
\end{align}
The upper bound on the scale of spontaneous $CP$ violation is therefore close to the lower bound on new flavor violating physics from meson mixing~\cite{Flavour_con} $\Lambda_F \gtrsim 10^7$ GeV, which hints toward future detection possibilities. Moreover, recent low scale ultraviolet completions of the Standard Model could be realized at energy scales as low as $10^8$ GeV~\cite{LowScale1,LowScale2}.

\section{Nelson-Barr Mechanism in Trinification}
\label{NBTri}
The introduction of vectorlike down-type quarks $D$ and additional scalar doublets is \textit{ad hoc} from a bottom-up perspective. However, this particle content arises automatically in trinification~\cite{333-1,333-2,333-3,333-4,333-5,333-6}, a gauge theory based on
\begin{align}
G_{333} = SU(3)_C \times SU(3)_L \times SU(3)_R \,,
\end{align}
which is one possible ultraviolet completion of left-right symmetric theories. The Nelson-Barr mechanism in $E_6$ grand unified theories was recently discussed in Ref.~\cite{NB_GUT}. The fermion content per generation in trinification is given by
\begin{align}
Q_L \sim (3, \bar{3}, 1)\,,\quad  Q_R \sim (3, 1, \bar{3})\,,\quad L_L \sim (1, 3, \bar{3})\,,
\end{align}
where the additional down-type quarks are embedded in $Q$. To reproduce the Standard Model mass spectrum, at least two scalar bitriplets are required
\begin{align}
\Phi_i \sim (1, 3, \bar{3}) \,,
\end{align}
with $i \in \{1,2\}$. The appearance of at least two left- and right-handed Higgs doublets thus follows from the necessity to reproduce the Standard Model. The interactions take the form
\begin{align}
-\mathcal{L} \supset \, & \bar{Q}_L \left(Y_1 \Phi_1 + Y_2 \Phi_2 \right) Q_R + h_1 L_L L_L \Phi_1 + h_2 L_L L_L \Phi_2 + \text{H.c.} \,.
\end{align}
Trinification thus motivates the minimal particle content needed in left-right symmetric theories to solve the strong $CP$ problem via spontaneous $CP$ violation. Unfortunately, this minimal scalar sector requires tuning of the scalar potential to solve the strong $CP$ problem via spontaneously breaking $CP$. To illustrate the tuning, we choose the simplified potential
\begin{align}
V =& -\mu_1 \text{Tr}\left(\Phi_1^\dagger \Phi_1 \right) -\mu_2 \text{Tr}\left(\Phi_2^\dagger \Phi_2 \right) -\mu_{12} \text{Tr}\left(\Phi_1^\dagger \Phi_2 \right) + \lambda_1 \text{Tr}\left(\Phi_1^\dagger \Phi_1 \Phi_1^\dagger \Phi_1 \right) + \lambda_2 \text{Tr}\left(\Phi_2^\dagger \Phi_2 \Phi_2^\dagger \Phi_2 \right) \nonumber \\
&\, + \lambda_{11} \text{Tr}\left(\Phi_1^\dagger \Phi_1\right)\text{Tr}\left(\Phi_1^\dagger \Phi_1 \right) + \lambda_{22} \text{Tr}\left(\Phi_2^\dagger \Phi_2\right)\text{Tr}\left(\Phi_2^\dagger \Phi_2 \right) + \lambda_\Phi \text{Tr}\left(\Phi_1^\dagger \Phi_2 \Phi_1^\dagger \Phi_2 \right) + \text{H.c.} \,.
\end{align}
In order to have a vacuum structure of the form
\begin{align}
\langle \Phi_1 \rangle = \begin{pmatrix} v_{11} && 0 &&  v_{L1} \\ 0 && v_{21} && 0 \\ v_{R1} && 0 && V_{31} \end{pmatrix} \quad \text{ and } \quad \langle \Phi_2\rangle = \begin{pmatrix} v_{12} && 0 && v_{L2} \\ 0 && v_{22} && 0 \\ v_{R2}\, e^{i\alpha} && 0 && V_{32} \end{pmatrix} \,, 
\end{align}
with $v_{L1} = 0$ and $v_{L2} = 0$, which is a necessary condition to solve the strong $CP$ problem via spontaneous $CP$ violation, we have to require
\begin{align}
\lambda_1 = \frac{v_{12}^2 V_{32}^2}{v_{11}^2 V_{31}^2}\, \lambda_2 \quad \text{ and } \quad \lambda_\Phi = -\frac{v_{12} V_{32}}{v_{11} V_{31}}\, \lambda_2 \,.
\end{align}

In order to avoid such tuning, we have to introduce a scalar bisextet
\begin{align}
\Sigma \sim (1, 6, \bar{6}) \,,
\end{align}
such that the dimension five and six operators
\begin{align}
V \supset \frac{c_5}{\Lambda} \Phi \Sigma \Sigma \Sigma \Phi^\dagger + \frac{c_6}{\Lambda^2} \Phi \Phi \Sigma \Sigma \Sigma \Sigma \,,
\end{align}
can generate a complex left-right symmetry breaking vacuum expectation value without also introducing a complex $ SU(2)_L $ violating vacuum expectation value.

However, the bisextet $\Sigma$ also has the advantage of enabling a type-II seesaw mechanism in the context of trinification. In the minimal theory of trinification, Majorana neutrino masses are generated at the one-loop level~\cite{333-2,Ohmer,333-nu}. As the neutral component of the $SU(2)_R$ triplet embedded in the bisextet $\Sigma$ develops a vacuum expectation value which breaks $SU(2)_R$, the right-handed neutrinos receive a Majorana mass term via the interaction
\begin{align}
-\mathcal{L} \supset h_\Sigma L_L \Sigma L_L \,.
\end{align}
The active neutrino mass is then suppressed by the ratio of electroweak vacuum expectation values to left-right symmetry breaking vacuum expectation values.

\section{Phenomenology of the Nelson-Barr Mechanism}
\label{Pheno}
The clearest prediction of the Nelson-Barr mechanism are new vectorlike quarks. Current LHC constraints from Atlas and CMS on vectorlike quarks can be found in Ref.~\cite{VLQLHC00,VLQLHC0,VLQLHC1,VLQLHC2,VLQLHC3,VLQLHC4,VLQLHC5} and give an approximate lower bound
\begin{align}
M_D \gtrsim 800 \text{ GeV}\,.
\end{align}
Constraints on theories with vectorlike quarks and additional colored fields such as for example trinification were studied in Ref.~\cite{VLQ333}. Also the possibility to search for vectorlike quarks with a broad width at the LHC was recently discussed in Ref.~\cite{VLQ_LHC_LW}. Assuming $M_D \simeq v_R$, we expect that a large range of the allowed parameter space can be tested at future hadron colliders. For projected limits on vectorlike quarks at a $100$ TeV proton-proton collider see Ref.~\cite{vlQ_fcc}.

The lower limits from right-handed current searches give the lower limit on the mass of $W_R$~\cite{LRbound1,LRbound2,LRbound3,LRbound4}
\begin{align}
M_{W_R} \gtrsim 3 \text{ TeV} \,,
\end{align}
where the mass is related to the scale of left-right symmetry breaking by the relation
\begin{align}
M_{W_R} \simeq \frac{g_R}{2} v_R\,,
\end{align}
where $g_R$ is the gauge coupling constant of $SU(2)_R$. Thus, for order one coupling constants the lower bound on $v_R$ is in the TeV range. We can therefore conclude that the left-right symmetry breaking scale $v_R$ in a theory with spontaneous $CP$ violation is bounded from above and below
\begin{align}
10^3 \text{ GeV} \lesssim v_R \lesssim 10^8 \text{ GeV}\,.
\end{align}
Note that for left-right symmetric theories with complex electroweak symmetry breaking vacuum expectation values stronger constraints on $M_{W_R}$ from the kaon sector apply~\cite{complex_kaon}.

Baryogenesis requires $C$ and $CP$ violation. A possibility to explain the baryon asymmetry in the Universe with spontaneous $CP$ violation is electroweak baryogenesis. The prospects of a strong first order phase transition at the electroweak scale in the minimal left-right symmetric theory with spontaneous $CP$ violation will be discussed in future work. Note that electroweak baryogenesis is testable at future gravitational wave observatories.

Measuring the reheating temperature of the Universe $T_{\text{RH}}$ or the Hubble rate during inflation $H_\text{inf}$ to be above $\sim10^8$ GeV would immediately rule out this minimal theory of spontaneous $CP$ violation. By spontaneously breaking the discrete $P$ and $CP$ symmetry at the scale $v_R$, we introduce domain walls into the thermal history of the Universe. As no such topological defects have been observed, we have to require that the left-right symmetry was never in the unbroken phase after inflation. Low-scale standard single field inflation predicts the tensor-to-scalar ratio $r$ to be~\cite{bound_r}
\begin{align}
r \simeq 1.5 \cdot 10^{-13} \left(\frac{H_\text{inf}}{10^8 \text{GeV}}\right)^2 \,.
\end{align}
The upper bound on the Hubble rate during inflation
\begin{align}
H_\text{inf} \lesssim 10^8 \text{ GeV}\,,
\end{align}
thus translates to an upper bound on the tensor-to-scalar ration 
\begin{align}
r \lesssim 10^{-13}\,. 
\end{align}
Therefore, if a sizeable tensor-to-scalar ratio $r$ is measured, the presented model of spontaneous $CP$ violation is ruled out.

The minimal Nelson-Barr mechanism in left-right symmetric theories presented in Sec. \ref{NBLR} predicts no $CP$ violation in the leptonic sector. This prediction is currently in tension with global fits to current neutrino data~\cite{globalFit1,globalFit2,globalFit3}. Future neutrino experiments such as DUNE~\cite{DUNE} and Hyper-Kamiokande~\cite{HyperK} will therefore be able to conclusively probe the minimal Nelson-Barr mechanism in left-right symmetric theories.

There are two possibilities to account for $CP$ violation in the leptonic sector at the renormalizable level.
\begin{itemize}
\item We can add a gauge singlet fermion
\begin{align}
N_L \sim (1, 1, 1, 0)\,.
\end{align}
The interactions
\begin{align}
-\mathcal{L} \supset y^{(N)}_{R,i} \bar{\ell}_R H^*_{R,i} N_L + m_N N^c_L N_L + \text{H.c.} \,,
\end{align}
lead to a mixing of the right-handed neutrinos $\nu_R$ and the Majorana fermion $N_L$. The complex vacuum expectation value $\langle H_{R,1} \rangle$ therefore enters the neutrino mass matrix. The gauge singlet $N_L$ is a potential dark matter candidate and can generate the baryon asymmetry in the early Universe via oscillations~\cite{BAU_osc}.
\item We can include new charged vectorlike leptons
\begin{align}
l_L \sim (1, 1, 1, -2) \,, \quad l_R \sim (1, 1, 1, -2)\,,
\end{align}
which mix via the interaction
\begin{align}
-\mathcal{L} \supset y^{(l)}_{L,i} \bar{\ell}_L H_{L,i} l_R +  y^{(l)}_{R,i} \bar{\ell}_R H_{R,i} l_L + m_L \bar{l}_L l_R + \text{H.c.} \,,
\end{align}
with the Standard Model leptons. The complex phase which spontaneously breaks $CP$ is thus introduced to the charged lepton mass matrix. 
\end{itemize}
In both scenarios the $CP$ violation in the quark sector and in the leptonic sector have a common origin. New vectorlike leptons up to masses of $100$ TeV can be probed with future experiments which measure the electric dipole moment of the electron, the rate of $\mu \to 3e$, and the rate of $\mu \to e\gamma$~\cite{leptons_lowE}. Additionally, vectorlike leptons are a striking signature at future lepton colliders, projected limits can be found in Ref.~\cite{vlL_fcc}. We thus see the complementarity of future neutrino experiments, low energy flavor experiments, and lepton and hadron colliders.

\section{Conclusions}
\label{con}
In this paper, we reevaluated the possibility of solving the strong $CP$ problem by spontaneous $CP$ violation in left-right symmetric theories. We therefore constructed in Sec. \ref{NBLR} the left-right symmetric theory with the minimal number of propagating degrees of freedom which can realize the Nelson-Barr mechanism. We add vectorlike down-type quarks, two left- and right-handed Higgs doublets and a scalar bidoublet to the Standard Model fermion content with three right-handed neutrinos. The $CP$ violation in the Standard Model is then directly related to the complex phase of the vacuum expectation value of a right-handed Higgs doublet as shown in Eq. (\ref{eq_relphase}). The QCD vacuum angle $\theta_{\text{QCD}}$ is only generated at loop-level. Interestingly, we can derive an upper bound on the scale of spontaneous $CP$ violation by taking into account the dimension five operator $H_{R,1}^\dagger H_{R,2} \bar{D}D + 1 \leftrightarrow 2$. As a consequence, the theory predicts that the reheating temperature after inflation and the Hubble rate during inflation have to be below $\sim 10^8$ GeV. We can therefore derive an upper bound on the tensor-to-scalar ratio $r \lesssim 10^{-13}$ such that this theory is testable.

The particle content of the minimal bottom-up approach to spontaneous $CP$ violation in left-right symmetric theories is well-motivated in the context of trinification. Moreover, trinification is an ultraviolet completion of the Standard Model which can be realized at scales as low as $\sim 10^8$ GeV. Note that a scalar bisextet $\Sigma$ is needed to break $CP$ spontaneously in trinification. However, the scalar bisextet also opens up the possibility of a type II seesaw mechanism in trinification.

The minimal model predicts no leptonic $CP$ violation. In the near future, current and up-coming neutrino experiments can therefore falsify the minimal model. To account for leptonic $CP$ violation new fermions have to be introduced. We presented the two options, including a gauge singlet fermion or additional vectorlike charged leptons, to incorporate the $CP$ violating phase in the leptonic sector at the renormalizable level. 

We can thus conclude that spontaneous $CP$ violation is still a theoretically motivated solution to the strong $CP$ problem. The minimal model of spontaneous $CP$ violation in left-right symmetric theories is highly predictive. Determining the Hubble rate during inflation and measuring the $CP$ phase in the neutrino mixing matrix can both potentially rule out the presented model.

\section{Acknowledgment} 
We thank Pavel Fileviez P\'erez, Jisuke Kubo, Ravi Kuchimanchi, Manfred Lindner, and Kai Schmitz for very useful discussions and comments on the manuscript.

\bibliography{ref_sCPV}{}

\providecommand{\href}[2]{#2}\begingroup\raggedright\begin{thebibliography}{10}

\bibitem{tHooft1}
G.~'t~Hooft, ``{\em {Symmetry Breaking Through Bell-Jackiw Anomalies}},''
\href{http://dx.doi.org/10.1103/PhysRevLett.37.8}{Phys. Rev. Lett. {\bf 37}
  (1976)  8--11}.

\bibitem{tHooft2}
G.~'t~Hooft, ``{\em {Computation of the Quantum Effects Due to a
  Four-Dimensional Pseudoparticle}},''
  \href{http://dx.doi.org/10.1103/PhysRevD.18.2199.3,
  10.1103/PhysRevD.14.3432}{Phys. Rev. {\bf D14} (1976)  3432--3450}.
[Erratum: Phys. Rev.D18,2199(1978)].

\bibitem{EDM_bound}
C.~A. Baker {\em et al.}, ``{\em {An Improved experimental limit on the
  electric dipole moment of the neutron}},''
  \href{http://dx.doi.org/10.1103/PhysRevLett.97.131801}{Phys. Rev. Lett. {\bf
  97} (2006)  131801},
\href{http://arxiv.org/abs/hep-ex/0602020}{{\tt arXiv:hep-ex/0602020}}.

\bibitem{EDM_mercury1}
B.~Graner, Y.~Chen, E.~G. Lindahl, and B.~R. Heckel, ``{\em {Reduced Limit on
  the Permanent Electric Dipole Moment of Hg199}},''
  \href{http://dx.doi.org/10.1103/PhysRevLett.119.119901,
  10.1103/PhysRevLett.116.161601}{Phys. Rev. Lett. {\bf 116} (2016) no.~16,
  161601}, \href{http://arxiv.org/abs/1601.04339}{{\tt arXiv:1601.04339}}.
[Erratum: Phys. Rev. Lett.119,no.11,119901(2017)].

\bibitem{EDM_mercury2}
N.~Yamanaka, B.~K. Sahoo, N.~Yoshinaga, T.~Sato, K.~Asahi, and B.~P. Das,
  ``{\em {Probing exotic phenomena at the interface of nuclear and particle
  physics with the electric dipole moments of diamagnetic atoms: A unique
  window to hadronic and semi-leptonic CP violation}},''
  \href{http://dx.doi.org/10.1140/epja/i2017-12237-2}{Eur. Phys. J. {\bf A53}
  (2017) no.~3, 54},
\href{http://arxiv.org/abs/1703.01570}{{\tt arXiv:1703.01570}}.

\bibitem{Solution_mU0_1}
D.~B. Kaplan and A.~V. Manohar, ``{\em {Current Mass Ratios of the Light
  Quarks}},''
\href{http://dx.doi.org/10.1103/PhysRevLett.56.2004}{Phys. Rev. Lett. {\bf 56}
  (1986)  2004}.

\bibitem{Solution_mU0_2}
T.~Banks, Y.~Nir, and N.~Seiberg, ``{\em {Missing (up) mass, accidental
  anomalous symmetries, and the strong CP problem}},'' in {\em {Yukawa
  couplings and the origins of mass. Proceedings, 2nd IFT Workshop,
  Gainesville, USA, February 11-13, 1994}}, pp.~26--41.
\newblock
\href{http://arxiv.org/abs/hep-ph/9403203}{{\tt arXiv:hep-ph/9403203}}.
\newblock

\bibitem{PQ1}
R.~D. Peccei and H.~R. Quinn, ``{\em {CP Conservation in the Presence of
  Instantons}},''
\href{http://dx.doi.org/10.1103/PhysRevLett.38.1440}{Phys. Rev. Lett. {\bf 38}
  (1977)  1440--1443}.

\bibitem{PQ2}
R.~D. Peccei and H.~R. Quinn, ``{\em {Constraints Imposed by CP Conservation in
  the Presence of Instantons}},''
\href{http://dx.doi.org/10.1103/PhysRevD.16.1791}{Phys. Rev. {\bf D16} (1977)
  1791--1797}.

\bibitem{sCPV1}
S.~M. Barr, ``{\em {A Natural Class of Nonpeccei-quinn Models}},''
\href{http://dx.doi.org/10.1103/PhysRevD.30.1805}{Phys. Rev. {\bf D30} (1984)
  1805}.

\bibitem{sCPV2}
S.~M. Barr, ``{\em {A Survey of a New Class of Models of {CP} Violation}},''
\href{http://dx.doi.org/10.1103/PhysRevD.34.1567}{Phys. Rev. {\bf D34} (1986)
  1567}.

\bibitem{sCPV3}
A.~E. Nelson, ``{\em {Naturally Weak CP Violation}},''
\href{http://dx.doi.org/10.1016/0370-2693(84)92025-2}{Phys. Lett. {\bf 136B}
  (1984)  387--391}.

\bibitem{sCPV4}
A.~E. Nelson, ``{\em {Calculation of $\theta$ Barr}},''
\href{http://dx.doi.org/10.1016/0370-2693(84)90827-X}{Phys. Lett. {\bf 143B}
  (1984)  165--170}.

\bibitem{Lattice_Review}
S.~Aoki {\em et al.}, ``{\em {Review of lattice results concerning low-energy
  particle physics}},''
  \href{http://dx.doi.org/10.1140/epjc/s10052-016-4509-7}{Eur. Phys. J. {\bf
  C77} (2017) no.~2, 112},
\href{http://arxiv.org/abs/1607.00299}{{\tt arXiv:1607.00299}}.

\bibitem{axion_review1}
J.~E. Kim and G.~Carosi, ``{\em {Axions and the Strong CP Problem}},''
  \href{http://dx.doi.org/10.1103/RevModPhys.82.557}{Rev. Mod. Phys. {\bf 82}
  (2010)  557--602},
\href{http://arxiv.org/abs/0807.3125}{{\tt arXiv:0807.3125}}.

\bibitem{axion_review2}
M.~Kawasaki and K.~Nakayama, ``{\em {Axions: Theory and Cosmological Role}},''
  \href{http://dx.doi.org/10.1146/annurev-nucl-102212-170536}{Ann. Rev. Nucl.
  Part. Sci. {\bf 63} (2013)  69--95},
\href{http://arxiv.org/abs/1301.1123}{{\tt arXiv:1301.1123}}.

\bibitem{axion_exp}
P.~W. Graham, I.~G. Irastorza, S.~K. Lamoreaux, A.~Lindner, and K.~A. van
  Bibber, ``{\em {Experimental Searches for the Axion and Axion-Like
  Particles}},''
  \href{http://dx.doi.org/10.1146/annurev-nucl-102014-022120}{Ann. Rev. Nucl.
  Part. Sci. {\bf 65} (2015)  485--514},
\href{http://arxiv.org/abs/1602.00039}{{\tt arXiv:1602.00039}}.

\bibitem{axion_exp2}
I.~G. Irastorza and J.~Redondo, ``{\em {New experimental approaches in the
  search for axion-like particles}},''
  \href{http://dx.doi.org/10.1016/j.ppnp.2018.05.003}{Prog. Part. Nucl. Phys.
  {\bf 102} (2018)  89--159},
\href{http://arxiv.org/abs/1801.08127}{{\tt arXiv:1801.08127}}.

\bibitem{axion_prob1}
M.~Kamionkowski and J.~March-Russell, ``{\em {Planck scale physics and the
  Peccei-Quinn mechanism}},''
  \href{http://dx.doi.org/10.1016/0370-2693(92)90492-M}{Phys. Lett. {\bf B282}
  (1992)  137--141},
\href{http://arxiv.org/abs/hep-th/9202003}{{\tt arXiv:hep-th/9202003}}.

\bibitem{axion_prob2}
R.~Holman, S.~D.~H. Hsu, T.~W. Kephart, E.~W. Kolb, R.~Watkins, and L.~M.
  Widrow, ``{\em {Solutions to the strong CP problem in a world with
  gravity}},'' \href{http://dx.doi.org/10.1016/0370-2693(92)90491-L}{Phys.
  Lett. {\bf B282} (1992)  132--136},
\href{http://arxiv.org/abs/hep-ph/9203206}{{\tt arXiv:hep-ph/9203206}}.

\bibitem{axion_prob3}
T.~Banks and N.~Seiberg, ``{\em {Symmetries and Strings in Field Theory and
  Gravity}},'' \href{http://dx.doi.org/10.1103/PhysRevD.83.084019}{Phys. Rev.
  {\bf D83} (2011)  084019},
\href{http://arxiv.org/abs/1011.5120}{{\tt arXiv:1011.5120}}.

\bibitem{NB_prob}
M.~Dine and P.~Draper, ``{\em {Challenges for the Nelson-Barr Mechanism}},''
  \href{http://dx.doi.org/10.1007/JHEP08(2015)132}{JHEP {\bf 08} (2015)  132},
\href{http://arxiv.org/abs/1506.05433}{{\tt arXiv:1506.05433}}.

\bibitem{BBP_model0}
L.~Bento and G.~C. Branco, ``{\em {Generation of a K-M phase from spontaneous
  CP breaking at a high-energy scale}},''
\href{http://dx.doi.org/10.1016/0370-2693(90)90697-5}{Phys. Lett. {\bf B245}
  (1990)  599--604}.

\bibitem{BBP_model}
L.~Bento, G.~C. Branco, and P.~A. Parada, ``{\em {A Minimal model with natural
  suppression of strong CP violation}},''
\href{http://dx.doi.org/10.1016/0370-2693(91)90530-4}{Phys. Lett. {\bf B267}
  (1991)  95--99}.

\bibitem{LR1}
J.~C. Pati and A.~Salam, ``{\em {Lepton Number as the Fourth Color}},''
  \href{http://dx.doi.org/10.1103/PhysRevD.10.275,
  10.1103/PhysRevD.11.703.2}{Phys. Rev. {\bf D10} (1974)  275--289}.
[Erratum: Phys. Rev.D11,703(1975)].

\bibitem{LR2}
R.~N. Mohapatra and J.~C. Pati, ``{\em A Natural Left-Right Symmetry},''
  \href{http://dx.doi.org/doi:10.1103/PhysRevD.11.2558}{Phys. Rev. D {\bf 11}
  (1975)  2558}.

\bibitem{LR3}
G.~Senjanovic, ``{\em Spontaneous Breakdown of Parity in a Class of Gauge
  Theories},'' \href{http://dx.doi.org/doi:10.1016/0550-3213(79)90604-7}{Nucl.
  Phys. B {\bf 153} (1979)  334}.

\bibitem{LR4}
G.~Senjanovic and R.~N. Mohapatra, ``{\em Exact Left-Right Symmetry and
  Spontaneous Violation of Parity},''
  \href{http://dx.doi.org/doi:10.1103/PhysRevD.12.1502}{Phys. Rev. D {\bf 12}
  (1975)  1502}.

\bibitem{Pconservation1}
M.~A.~B. Beg and H.~S. Tsao, ``{\em {Strong P, T Noninvariances in a Superweak
  Theory}},''
\href{http://dx.doi.org/10.1103/PhysRevLett.41.278}{Phys. Rev. Lett. {\bf 41}
  (1978)  278}.

\bibitem{Pconservation2}
R.~N. Mohapatra and G.~Senjanovic, ``{\em {Natural Suppression of Strong p and
  t Noninvariance}},''
\href{http://dx.doi.org/10.1016/0370-2693(78)90243-5}{Phys. Lett. {\bf 79B}
  (1978)  283--286}.

\bibitem{Hall}
L.~J. Hall and K.~Harigaya, ``{\em {Implications of Higgs Discovery for the
  Strong CP Problem and Unification}},''
\href{http://arxiv.org/abs/1803.08119}{{\tt arXiv:1803.08119}}.

\bibitem{LR_sCPV1}
K.~S. Babu and R.~N. Mohapatra, ``{\em {A Solution to the Strong {CP} Problem
  Without an Axion}},''
\href{http://dx.doi.org/10.1103/PhysRevD.41.1286}{Phys. Rev. {\bf D41} (1990)
  1286}.

\bibitem{LR_sCPV2}
R.~N. Mohapatra, A.~Rasin, and G.~Senjanovic, ``{\em {P, C and strong CP in
  left-right supersymmetric models}},''
  \href{http://dx.doi.org/10.1103/PhysRevLett.79.4744}{Phys. Rev. Lett. {\bf
  79} (1997)  4744--4747},
\href{http://arxiv.org/abs/hep-ph/9707281}{{\tt arXiv:hep-ph/9707281}}.

\bibitem{LR_CP1}
R.~N. Mohapatra and G.~Senjanovic, ``{\em {Natural Suppression of Strong p and
  t Noninvariance}},''
\href{http://dx.doi.org/10.1016/0370-2693(78)90243-5}{Phys. Lett. {\bf 79B}
  (1978)  283--286}.

\bibitem{LR_CP2}
S.~M. Barr, D.~Chang, and G.~Senjanovic, ``{\em {Strong CP problem and
  parity}},''
\href{http://dx.doi.org/10.1103/PhysRevLett.67.2765}{Phys. Rev. Lett. {\bf 67}
  (1991)  2765--2768}.

\bibitem{LR_CP3}
R.~Kuchimanchi, ``{\em {P/CP Conserving CP/P Violation Solves Strong CP
  Problem}},'' \href{http://dx.doi.org/10.1103/PhysRevD.82.116008}{Phys. Rev.
  {\bf D82} (2010)  116008},
\href{http://arxiv.org/abs/1009.5961}{{\tt arXiv:1009.5961}}.

\bibitem{LR_CP4}
Y.~Zhang, H.~An, X.~Ji, and R.~N. Mohapatra, ``{\em {General CP Violation in
  Minimal Left-Right Symmetric Model and Constraints on the Right-Handed
  Scale}},'' \href{http://dx.doi.org/10.1016/j.nuclphysb.2008.05.019}{Nucl.
  Phys. {\bf B802} (2008)  247--279},
\href{http://arxiv.org/abs/0712.4218}{{\tt arXiv:0712.4218}}.

\bibitem{LRspontaneousCPtriplets}
G.~Barenboim and J.~Bernabeu, ``{\em {Spontaneous breakdown of CP in left-right
  symmetric models}},'' \href{http://dx.doi.org/10.1007/s002880050321}{Z. Phys.
  {\bf C73} (1997)  321--331},
\href{http://arxiv.org/abs/hep-ph/9603379}{{\tt arXiv:hep-ph/9603379}}.

\bibitem{Nierste}
G.~Barenboim, M.~Gorbahn, U.~Nierste, and M.~Raidal, ``{\em {Higgs sector of
  the minimal left-right symmetric model}},''
  \href{http://dx.doi.org/10.1103/PhysRevD.65.095003}{Phys. Rev. {\bf D65}
  (2002)  095003},
\href{http://arxiv.org/abs/hep-ph/0107121}{{\tt arXiv:hep-ph/0107121}}.

\bibitem{sCPV_Cond}
H.~E. Haber and Z.~Surujon, ``{\em {A Group-theoretic Condition for Spontaneous
  CP Violation}},'' \href{http://dx.doi.org/10.1103/PhysRevD.86.075007}{Phys.
  Rev. {\bf D86} (2012)  075007},
\href{http://arxiv.org/abs/1201.1730}{{\tt arXiv:1201.1730}}.

\bibitem{minLR}
P.~Fileviez~Perez, C.~Murgui, and S.~Ohmer, ``{\em Simple Left-Right Theory:
  Lepton Number Violation at the LHC},''
  \href{http://dx.doi.org/10.1103/PhysRevD.94.051701}{Phys. Rev. {\bf D94}
  (2016) no.~5, 051701},
\href{http://arxiv.org/abs/1607.00246}{{\tt arXiv:1607.00246}}.

\bibitem{PDG}
{\bf Particle Data Group}, C.~Patrignani {\em et al.}, ``{\em {Review of
  Particle Physics}},''
\href{http://dx.doi.org/10.1088/1674-1137/40/10/100001}{Chin. Phys. {\bf C40}
  (2016) no.~10, 100001}.

\bibitem{nEDM1}
S.~K. Lamoreaux and R.~Golub, ``{\em {Experimental searches for the neutron
  electric dipole moment}},''
\href{http://dx.doi.org/10.1088/0954-3899/36/10/104002}{J. Phys. {\bf G36}
  (2009)  104002}.

\bibitem{nEDM2}
C.~A. Baker {\em et al.}, ``{\em {The search for the neutron electric dipole
  moment at the Paul Scherrer Institute}},''
\href{http://dx.doi.org/10.1016/j.phpro.2011.06.032}{Phys. Procedia {\bf 17}
  (2011)  159--167}.

\bibitem{nEDM3}
{\bf nEDM}, E.~P. Tsentalovich, ``{\em {The nEDM experiment at the SNS}},''
\href{http://dx.doi.org/10.1134/S1063779614011073}{Phys. Part. Nucl. {\bf 45}
  (2014)  249--250}.

\bibitem{Flavour_con}
G.~Isidori and F.~Teubert, ``{\em {Status of indirect searches for New Physics
  with heavy flavour decays after the initial LHC run}},''
  \href{http://dx.doi.org/10.1140/epjp/i2014-14040-4}{Eur. Phys. J. Plus {\bf
  129} (2014) no.~3, 40},
\href{http://arxiv.org/abs/1402.2844}{{\tt arXiv:1402.2844}}.

\bibitem{LowScale1}
P.~Fileviez~Perez and S.~Ohmer, ``{\em {Low Scale Unification of Gauge
  Interactions}},'' \href{http://dx.doi.org/10.1103/PhysRevD.90.037701}{Phys.
  Rev. {\bf D90} (2014) no.~3, 037701},
\href{http://arxiv.org/abs/1405.1199}{{\tt arXiv:1405.1199}}.

\bibitem{LowScale2}
P.~Fileviez~Perez and S.~Ohmer, ``{\em {Unification and Local Baryon
  Number}},'' \href{http://dx.doi.org/10.1016/j.physletb.2017.02.049}{Phys.
  Lett. {\bf B768} (2017)  86--91},
\href{http://arxiv.org/abs/1612.07165}{{\tt arXiv:1612.07165}}.

\bibitem{333-1}
K.~S. Babu, X.~G. He, and S.~Pakvasa, ``{\em Neutrino Masses and Proton Decay
  Modes in SU(3) X SU(3) X SU(3) Trinification},''
  \href{http://dx.doi.org/doi:10.1103/PhysRevD.33.763}{Phys. Rev. D {\bf 33}
  (1986)  763}.

\bibitem{333-2}
J.~Sayre, S.~Wiesenfeldt, and S.~Willenbrock, ``{\em Minimal trinification},''
  \href{http://dx.doi.org/doi:10.1103/PhysRevD.73.035013}{Phys. Rev. D {\bf 73}
  (2006)  035013}, \href{http://arxiv.org/abs/hep-ph/0601040}{{\tt
  hep-ph/0601040}}.

\bibitem{333-3}
J.~Hetzel and B.~Stech, ``{\em Low-energy phenomenology of trinification: an
  effective left-right-symmetric model},''
  \href{http://dx.doi.org/doi:10.1103/PhysRevD.91.055026}{Phys. Rev. D {\bf 91}
  (2015)  055026}, \href{http://arxiv.org/abs/1502.00919}{{\tt
  arXiv:1502.00919}}.

\bibitem{333-4}
J.~Hetzel, ``{\em Phenomenology of a left-right-symmetric model inspired by the
  trinification model},'' \href{http://arxiv.org/abs/1504.06739}{{\tt
  arXiv:1504.06739}}.

\bibitem{333-5}
G.~M. Pelaggi, A.~Strumia, and E.~Vigiani, ``{\em Trinification can explain the
  di-photon and di-boson LHC anomalies},''
  \href{http://dx.doi.org/doi:10.1007/JHEP03(2016)025}{JHEP {\bf 1603} (2016)
  025}, \href{http://arxiv.org/abs/1512.07225}{{\tt arXiv:1512.07225}}.

\bibitem{333-6}
G.~M. Pelaggi, A.~Strumia, and S.~Vignali, ``{\em Totally asymptotically free
  trinification},'' \href{http://dx.doi.org/doi:10.1007/JHEP08(2015)130}{JHEP
  {\bf 1508} (2015)  130}, \href{http://arxiv.org/abs/1507.06848}{{\tt
  arXiv:1507.06848}}.

\bibitem{NB_GUT}
J.~Schwichtenberg, P.~Tremper, and R.~Ziegler, ``{\em {A Grand-Unified
  Nelson-Barr Model}},''
\href{http://arxiv.org/abs/1802.08109}{{\tt arXiv:1802.08109}}.

\bibitem{Ohmer}
S.~Ohmer, \href{http://dx.doi.org/10.11588/heidok.00023860,
  urn:nbn:de:bsz:16-heidok-238600}{{\em {New Symmetry Concepts for Spacetime
  and Unification}}}.
\newblock PhD thesis, U. Heidelberg (main), 2017.
\newblock
\url{http://archiv.ub.uni-heidelberg.de/archiv/23860/}.
\newblock

\bibitem{333-nu}
K.~S. Babu, B.~Bajc, M.~Nemevšek, and Z.~Tavartkiladze, ``{\em {Trinification
  at the TeV scale}},''
\href{http://dx.doi.org/10.1063/1.5010106}{AIP Conf. Proc. {\bf 1900} (2017)
  no.~1, 020002}.

\bibitem{VLQLHC00}
{\bf ATLAS}, G.~Aad {\em et al.}, ``{\em {Search for pair and single production
  of new heavy quarks that decay to a $Z$ boson and a third-generation quark in
  $pp$ collisions at $\sqrt{s}=8$ TeV with the ATLAS detector}},''
  \href{http://dx.doi.org/10.1007/JHEP11(2014)104}{JHEP {\bf 11} (2014)  104},
\href{http://arxiv.org/abs/1409.5500}{{\tt arXiv:1409.5500}}.

\bibitem{VLQLHC0}
{\bf ATLAS}, G.~Aad {\em et al.}, ``{\em {Search for vector-like $B$ quarks in
  events with one isolated lepton, missing transverse momentum and jets at
  $\sqrt{s}=$ 8 TeV with the ATLAS detector}},''
  \href{http://dx.doi.org/10.1103/PhysRevD.91.112011}{Phys. Rev. {\bf D91}
  (2015) no.~11, 112011},
\href{http://arxiv.org/abs/1503.05425}{{\tt arXiv:1503.05425}}.

\bibitem{VLQLHC1}
{\bf CMS}, V.~Khachatryan {\em et al.}, ``{\em {Search for pair-produced
  vectorlike B quarks in proton-proton collisions at $\sqrt{s}$=8  TeV}},''
  \href{http://dx.doi.org/10.1103/PhysRevD.93.112009}{Phys. Rev. {\bf D93}
  (2016) no.~11, 112009},
\href{http://arxiv.org/abs/1507.07129}{{\tt arXiv:1507.07129}}.

\bibitem{VLQLHC2}
{\bf ATLAS}, G.~Aad {\em et al.}, ``{\em {Search for pair production of a new
  heavy quark that decays into a $W$ boson and a light quark in $pp$ collisions
  at $\sqrt{s} = 8$ TeV with the ATLAS detector}},''
  \href{http://dx.doi.org/10.1103/PhysRevD.92.112007}{Phys. Rev. {\bf D92}
  (2015) no.~11, 112007},
\href{http://arxiv.org/abs/1509.04261}{{\tt arXiv:1509.04261}}.

\bibitem{VLQLHC3}
{\bf CMS}, A.~M. Sirunyan {\em et al.}, ``{\em {Search for single production of
  vector-like quarks decaying to a Z boson and a top or a bottom quark in
  proton-proton collisions at $ \sqrt{s}=13 $ TeV}},''
  \href{http://dx.doi.org/10.1007/JHEP05(2017)029}{JHEP {\bf 05} (2017)  029},
\href{http://arxiv.org/abs/1701.07409}{{\tt arXiv:1701.07409}}.

\bibitem{VLQLHC4}
{\bf ATLAS}, M.~Aaboud {\em et al.}, ``{\em {Search for pair production of
  heavy vector-like quarks decaying to high-p$_{T}$ W bosons and b quarks in
  the lepton-plus-jets final state in pp collisions at $ \sqrt{s}=13 $ TeV with
  the ATLAS detector}},'' \href{http://dx.doi.org/10.1007/JHEP10(2017)141}{JHEP
  {\bf 10} (2017)  141},
\href{http://arxiv.org/abs/1707.03347}{{\tt arXiv:1707.03347}}.

\bibitem{VLQLHC5}
{\bf CMS}, A.~M. Sirunyan {\em et al.}, ``{\em {Search for vectorlike
  light-flavor quark partners in proton-proton collisions at $\sqrt s$
  =8  TeV}},'' \href{http://dx.doi.org/10.1103/PhysRevD.97.072008}{Phys.
  Rev. {\bf D97} (2018)  072008},
\href{http://arxiv.org/abs/1708.02510}{{\tt arXiv:1708.02510}}.

\bibitem{VLQ333}
A.~Deandrea and A.~M. Iyer, ``{\em {Vectorlike quarks and heavy colored bosons
  at the LHC}},'' \href{http://dx.doi.org/10.1103/PhysRevD.97.055002}{Phys.
  Rev. {\bf D97} (2018) no.~5, 055002},
\href{http://arxiv.org/abs/1710.01515}{{\tt arXiv:1710.01515}}.

\bibitem{VLQ_LHC_LW}
A.~Carvalho, S.~Moretti, D.~O'Brien, L.~Panizzi, and H.~Prager, ``{\em {Single
  production of vector-like quarks with large width at the Large Hadron
  Collider}},''
\href{http://arxiv.org/abs/1805.06402}{{\tt arXiv:1805.06402}}.

\bibitem{vlQ_fcc}
M.~Chala, R.~Gröber, and M.~Spannowsky, ``{\em {Searches for vector-like
  quarks at future colliders and implications for composite Higgs models with
  dark matter}},'' \href{http://dx.doi.org/10.1007/JHEP03(2018)040}{JHEP {\bf
  03} (2018)  040},
\href{http://arxiv.org/abs/1801.06537}{{\tt arXiv:1801.06537}}.

\bibitem{LRbound1}
A.~Maiezza, M.~Nemevsek, F.~Nesti, and G.~Senjanovic, ``{\em {Left-Right
  Symmetry at LHC}},''
  \href{http://dx.doi.org/10.1103/PhysRevD.82.055022}{Phys. Rev. {\bf D82}
  (2010)  055022},
\href{http://arxiv.org/abs/1005.5160}{{\tt arXiv:1005.5160}}.

\bibitem{LRbound2}
Y.~Zhang, H.~An, X.~Ji, and R.~N. Mohapatra, ``{\em General CP Violation in
  Minimal Left-Right Symmetric Model and Constraints on the Right-Handed
  Scale},'' \href{http://dx.doi.org/doi:10.1016/j.nuclphysb.2008.05.019}{Nucl.
  Phys. B {\bf 802} (2008)  247}, \href{http://arxiv.org/abs/0712.4218}{{\tt
  arXiv:0712.4218}}.

\bibitem{LRbound3}
M.~Blanke, A.~J. Buras, K.~Gemmler, and T.~Heidsieck, ``{\em Delta $F = 2$
  observables and $B \to X_q \gamma$ decays in the Left-Right Model: Higgs
  particles striking back},''
  \href{http://dx.doi.org/doi:10.1007/JHEP03(2012)024}{JHEP {\bf 1203} (2012)
  024}, \href{http://arxiv.org/abs/1111.5014}{{\tt arXiv:1111.5014}}.

\bibitem{LRbound4}
S.~Bertolini, A.~Maiezza, and F.~Nesti, ``{\em Present and Future K and B Meson
  Mixing Constraints on TeV Scale Left-Right Symmetry},''
  \href{http://dx.doi.org/doi:10.1103/PhysRevD.89.095028}{Phys. Rev. D {\bf 89}
  (2014)  no.9, 095028}, \href{http://arxiv.org/abs/1403.7112}{{\tt
  arXiv:1403.7112}}.

\bibitem{complex_kaon}
A.~Maiezza and M.~Nemevsek, ``{\em {Strong P invariance, neutron electric
  dipole moment, and minimal left-right parity at LHC}},''
  \href{http://dx.doi.org/10.1103/PhysRevD.90.095002}{Phys. Rev. {\bf D90}
  (2014) no.~9, 095002},
\href{http://arxiv.org/abs/1407.3678}{{\tt arXiv:1407.3678}}.

\bibitem{bound_r}
K.~Schmitz and T.~T. Yanagida, ``{\em {Axion Isocurvature Perturbations in
  Low-Scale Models of Hybrid Inflation}},''
\href{http://arxiv.org/abs/1806.06056}{{\tt arXiv:1806.06056}}.

\bibitem{globalFit1}
I.~Esteban, M.~C. Gonzalez-Garcia, M.~Maltoni, I.~Martinez-Soler, and
  T.~Schwetz, ``{\em {Updated fit to three neutrino mixing: exploring the
  accelerator-reactor complementarity}},''
  \href{http://dx.doi.org/10.1007/JHEP01(2017)087}{JHEP {\bf 01} (2017)  087},
\href{http://arxiv.org/abs/1611.01514}{{\tt arXiv:1611.01514}}.

\bibitem{globalFit2}
P.~F. de~Salas, D.~V. Forero, C.~A. Ternes, M.~Tortola, and J.~W.~F. Valle,
  ``{\em {Status of neutrino oscillations 2018: first hint for normal mass
  ordering and improved CP sensitivity}},''
\href{http://arxiv.org/abs/1708.01186}{{\tt arXiv:1708.01186}}.

\bibitem{globalFit3}
F.~Capozzi, E.~Lisi, A.~Marrone, and A.~Palazzo, ``{\em {Current unknowns in
  the three neutrino framework}},''
\href{http://arxiv.org/abs/1804.09678}{{\tt arXiv:1804.09678}}.

\bibitem{DUNE}
{\bf DUNE}, R.~Acciarri {\em et al.}, ``{\em {Long-Baseline Neutrino Facility
  (LBNF) and Deep Underground Neutrino Experiment (DUNE)}},''
\href{http://arxiv.org/abs/1512.06148}{{\tt arXiv:1512.06148}}.

\bibitem{HyperK}
{\bf Hyper-Kamiokande}, K.~Abe {\em et al.}, ``{\em {Hyper-Kamiokande Design
  Report}},''
\href{http://arxiv.org/abs/1805.04163}{{\tt arXiv:1805.04163}}.

\bibitem{BAU_osc}
E.~K. Akhmedov, V.~A. Rubakov, and A.~{\relax Yu}. Smirnov, ``{\em
  {Baryogenesis via neutrino oscillations}},''
  \href{http://dx.doi.org/10.1103/PhysRevLett.81.1359}{Phys. Rev. Lett. {\bf
  81} (1998)  1359--1362},
\href{http://arxiv.org/abs/hep-ph/9803255}{{\tt arXiv:hep-ph/9803255}}.

\bibitem{leptons_lowE}
K.~Ishiwata and M.~B. Wise, ``{\em {Phenomenology of heavy vectorlike
  leptons}},'' \href{http://dx.doi.org/10.1103/PhysRevD.88.055009}{Phys. Rev.
  {\bf D88} (2013) no.~5, 055009},
\href{http://arxiv.org/abs/1307.1112}{{\tt arXiv:1307.1112}}.

\bibitem{vlL_fcc}
E.~Bertuzzo, P.~A.~N. Machado, Y.~F. Perez-Gonzalez, and R.~Zukanovich~Funchal,
  ``{\em {Constraints from Triple Gauge Couplings on Vectorlike Leptons}},''
  \href{http://dx.doi.org/10.1103/PhysRevD.96.035035}{Phys. Rev. {\bf D96}
  (2017) no.~3, 035035},
\href{http://arxiv.org/abs/1706.03073}{{\tt arXiv:1706.03073}}.

\end{thebibliography}\endgroup
\bibliographystyle{utcaps_mod}

\end{document}